\documentclass[cameraready]{Interspeech}

\title{VoxEmo: Benchmarking Speech Emotion Recognition with Speech LLMs}
\author[affiliation={1},orcid=0000-0003-0122-990X]{Hezhao}{Zhang}
\author[affiliation={2},orcid=0000-0003-2125-5689]{Huang-Cheng}{Chou}
\author[affiliation={2},orcid=0000-0002-1052-6204]{Shrikanth}{Narayanan}
\author[affiliation={1}, orcid=0000-0003-0939-3464]{Thomas}{Hain}

\address{
    $^1$ Department of Computer Science, University of Sheffield, United Kingdom \\
    $^2$ Signal Analysis and Interpretation Laboratory (SAIL), Ming Hsieh Department of Electrical and Computer Engineering, University of Southern California, Los Angeles, CA 90089, USA 
}
\email{\{hzhang181, t.hain\}@sheffield.ac.uk}

\keywords{speech recognition, human-computer interaction, computational paralinguistics}

\usepackage{comment}
\usepackage{tabularx}
\usepackage{multirow}

\begin{document}

\maketitle

\begin{abstract}
Speech Large Language Models (LLMs) show great promise for speech emotion recognition (SER) via generative interfaces. 
However, shifting from closed-set classification to open text generation introduces zero-shot stochasticity, making evaluation highly sensitive to prompts. 
Additionally, conventional speech LLMs benchmarks overlook the inherent ambiguity of human emotion. 
Hence, we present VoxEmo, a comprehensive SER benchmark encompassing 35 emotion corpora across 15 languages for Speech LLMs. 
VoxEmo provides a standardized toolkit featuring varying prompt complexities, from direct classification to paralinguistic reasoning. 
To reflect real-world perception/application, we introduce a distribution-aware soft-label protocol and a prompt-ensemble strategy that emulates annotator disagreement. 
Experiments reveal that while zero-shot speech LLMs trail supervised baselines in hard-label accuracy, they uniquely align with human subjective distributions.
\end{abstract}

\section{Introduction}
Speech emotion recognition (SER) concerns the automatic recognition of expressed emotion and affect from speech \cite{schullerSpeechEmotionRecognition2018,lee2023engineering}. 
As a key capability for modeling affect in human-computer interaction (HCI), SER supports affect-aware spoken interfaces and speech analytics, such as spoken dialog systems \cite{Lee_2005} and call-center monitoring \cite{Gupta_2007}. 
Despite significant advancements in speech fields across multiple speech tasks in recent years (e.g., WavLM \cite{Chen_2022_wavlm}), achieving robust SER performance in naturalistic speech remains significantly more challenging than in controlled, acted settings \cite{Wu_2024}. 
Recently, a different strategy to solve the many challenges has emerged: SER that leverages the many abilities of speech Large Language Models (LLMs) \cite{Bellver_2024} (e.g., Qwen-Audio \cite{chuQwen2AudioTechnicalReport2024}). 
Speech LLMs have gained interest by achieving competitive results utilising generative, instruction-following interfaces \cite{maiAASLLMAcousticallyAugmented2025}, jointly leveraging lexical semantics and paralinguistic cues from speech \cite{cuiRecentAdvancesSpeech2025}.

However, the transition from traditional models to Speech LLMs does not relate well with existing evaluation paradigms. Traditionally, SER is formulated as a supervised, closed-set classification task \cite{schullerSpeechEmotionRecognition2018,yangSUPERBSpeechProcessing2021}. 
Typically, a speech encoder is used to extract embeddings, a lightweight prediction head outputs class probabilities, and inference-time behavior is strictly fixed by the model architecture \cite{pepinoEmotionRecognitionSpeech2021, Wu_2024}. 
Consequently, existing benchmark frameworks primarily focus on standardizing data splits and metrics \cite{maEmoBoxMultilingualMulticorpus2024,Wu_2024}. 
In contrast, speech LLMs based SER operates as a generative, text-output task. 
Given an audio clip and an explicit instruction/prompt, the model generates a text response that must be parsed and mapped to a closed label set \cite{murzakuOmniVoxZeroShotEmotion2025}. 
Reported scores are therefore highly sensitive to \textit{inference-time protocols}-, specifically prompt formulation, decoding settings, and parsing heuristics \cite{amin2024prompt}. 
This zero-shot stochasticity drastically reduces comparability across studies, necessitating evaluation procedures that make inference-time settings explicitly reproducible.

Beyond changes in how speech is modeled, SER benchmarking is confounded by the semantics of underlying datasets and domain shifts.  
Emotion is inherently ambiguous and subjective \cite{larrouy-maestriSoundEmotionalProsody2025,Chou_2025}, leading to low inter-annotator agreement, especially in naturalistic corpora \cite{lotfianBuildingNaturalisticEmotionally2019,bussoMSPPodcastCorpus2025}. 
This implies a crucial differentiation by \textit{Label Source}: whether annotations reflect the emotion intended/expressed by the speaker or the emotion perceived by the listeners \cite{carlosbussoExpressionPerceptionEmotions2008}. 
Prior work suggests that retaining label distributions (soft-labels) to reflect the annotation and label uncertainty is preferred to collapsing minority views into a single hard label \cite{chouMinorityViewsMatter2025,chouEmbracingAmbiguitySubjectivity2024}. 
Furthermore, as zero-shot speech LLMs enable seamless cross-corpus evaluation, unguided comparisons often conflate multiple variables, such as language, interaction style, or recording conditions \cite{milner2019cross,phukan2025rethinking}.  
Therefore, an appropriate benchmark should aim to categorise cross-corpus transfer performance using explicit, interpretable shift types.

To sum up, the effects of generative model stochasticity and dataset semantic ambiguity require a unified, large-scale benchmarking framework.  
We introduce VoxEmo, an evaluation toolkit and benchmark designed to provide a standard for investigations into speech LLM-based SER. 
Our contributions are 4-fold: 
\begin{itemize}
    \item (i) A standardized speech LLM-based SER evaluation toolkit that unifies prompt templates, generation settings, output parsing, and invalid-output handling.
    \item (ii) A comprehensive benchmark report and scoreboard spanning 35 corpora across 15 languages, accompanied by a disclosure checklist for reproducible comparisons.
    \item (iii) A novel dataset metadata schema that explicitly documents the \textbf{Label source} (perceived vs. expressed) to enable \textit{distribution-aware evaluation}.
    \item (iv) Structured, corpus-driven cross-domain settings that explicitly isolate label-set and acoustic shifts.
\end{itemize}

\begin{table*}[!t]
\centering
\begingroup
\fontsize{7}{9}\selectfont 
\setlength{\tabcolsep}{4pt}
\hyphenpenalty=10000\exhyphenpenalty=10000
\caption{\small Dataset overview, grouped by acted vs in-the-wild recordings and sorted by year. Audio source (\textbf{Aud. S.}) reports recording origin for in-the-wild corpora, and scripted, spontaneous, mixed (\textbf{Script/Spont}) style for acted corpora. Label source indicates available label sources (\textbf{Lab. S.}): Expr (expressed/actor-intended), Perc (perceived/listener-annotator). Emo, \#Utts, and \#Hrs reflect the benchmark subset after dataset-specific filtering and label-set restriction (e.g., excluding no-agreement samples in MSP-Podcast and using a commonly adopted label subset for IEMOCAP).}
\vspace{-3mm}
\begin{tabular*}{\textwidth}{@{\extracolsep{\fill}}>{\raggedright\arraybackslash}p{0.20\textwidth}>{\centering\arraybackslash}p{0.045\textwidth}>{\hspace{7pt}\raggedright\arraybackslash}p{0.145\textwidth}>{\raggedright\arraybackslash}p{0.085\textwidth}>{\centering\arraybackslash}p{0.045\textwidth}>{\raggedleft\arraybackslash}p{0.055\textwidth}>{\raggedleft\arraybackslash}p{0.105\textwidth}>{\raggedleft\arraybackslash}p{0.045\textwidth}>{\raggedright\arraybackslash}p{0.19\textwidth}@{}}
\toprule
\textbf{Dataset} & \textbf{Year} & \textbf{Aud. S.} & \textbf{Lab. S.} & \textbf{\#Emos} & \textbf{\#Spks} & \textbf{Selected \#Utts} & \textbf{\#Hrs} & \textbf{Langage} \\
\midrule
\multicolumn{9}{@{}c@{}}{\textbf{In-the-wild}} \\ \midrule
\addlinespace[0.1em]
EmotionTalk~\cite{sunEmotionTalkInteractiveChinese2025} & 2025 & Talkshow & Perc & 7 & 30 & 19,250 & 23.6 & Mandarin \\
MSP-Podcast~\cite{bussoMSPPodcastCorpus2025,lotfianBuildingNaturalisticEmotionally2019} & 2025 & Podcast & Perc & 8 & 3,711 & 212,801 & 324.1 & English \\
BIIC-Podcast~\cite{upadhyayIntelligentInfrastructureLarge2023} & 2023 & Podcast & Perc & 8 & -- & 69,874 & 147.2 & Taiwanese Mandarin \\
M3ED~\cite{zhaoM3EDMultimodalMultiscene2022} & 2022 & TV & Perc & 7 & 634 & 24,437 & 9.8 & Mandarin \\
ASVP-ESD~\cite{landry2020asvp} & 2020 & Online media & Perc & 14 & 132 & 13,964 & 18.0 & English/Mandarin \\
MELD~\cite{poriaMELDMultimodalMultiParty2019} & 2019 & TV & Perc & 7 & -- & 13,703 & 12.0 & English \\
URDU~\cite{latif2018cross} & 2018 & Talkshow & Perc & 4 & 38 & 400 & 0.3 & Urdu \\
\midrule
\addlinespace[0.1em]
\multicolumn{9}{@{}c@{}}{\textbf{Acted}} \\ \midrule
\addlinespace[0.1em]
EmoDB~\cite{burkhardtDatabaseGermanEmotional2005,burkhardtEmoDB20Database2025} & 2025 & Scripted & Expr & 7 & 10 & 816 & 0.6 & German \\
Thai-ser~\cite{wongpithayadisaiTHAISpeechEmotion2025} & 2025 & Script/Spont & Expr & 5 & 200 & 27,854 & 41.6 & Thai \\
nEMO~\cite{christop-2024-nemo} & 2024 & Scripted & Expr & 6 & 9 & 4,481 & 3.1 & Polish \\
ASED~\cite{retta2023new} & 2023 & Scripted & Perc & 5 & 65 & 2,474 & 2.1 & Amharic \\
EMNS~\cite{noriyEMNSImzCorpus2023} & 2023 & Scripted & Expr & 8 & 1 & 1,181 & 1.9 & English \\
Emozionalmente~\cite{catania2023speech} & 2023 & Scripted & Expr & 7 & 431 & 6,902 & 6.3 & Italian \\
RESD~\cite{aniemore2022resd} & 2022 & Unknown & Expr & 7 & -- & 1,337 & 2.1 & Russian \\
ESD~\cite{zhou2021seen} & 2021 & Scripted & Expr & 5 & 20 & 35,000 & 29.1 & English/Mandarin \\
MESD~\cite{duvilleMexicanEmotionalSpeech2021} & 2021 & Scripted & Expr & 6 & 16 & 862 & 0.2 & Spanish \\
MEAD~\cite{wangMEADLargeScaleAudioVisual2020} & 2020 & Scripted & Expr & 8 & 60 & 31,729 & 37.3 & English \\
Oreau~\cite{kerkeni2020french} & 2020 & Scripted & Expr & 7 & 32 & 434 & 0.3 & French \\
Polish~\cite{miesikowska2020emotions} & 2020 & Scripted & Expr & 3 & 5 & 450 & 0.1 & Polish \\
TurEV-DB~\cite{canpolatTurkishEmotionVoice2020} & 2020 & Scripted & Expr & 4 & 6 & 1,735 & 0.5 & Turkish \\
ShEMO~\cite{mohamadnezamiShEMOLargescaleValidated2019} & 2019 & Spontaneous & Perc & 6 & 87 & 2,838 & 3.3 & Persian \\
SUBESCO~\cite{sultanaSUSTBanglaEmotional2021} & 2019 & Scripted & Expr & 7 & 20 & 7,000 & 7.8 & Bangla \\
AESDD~\cite{vrysasSpeechEmotionRecognition2018} & 2018 & Scripted & Expr & 5 & 6 & 604 & 0.7 & Greek \\
CaFE~\cite{gournayCanadianFrenchEmotional2018} & 2018 & Scripted & Expr & 7 & 12 & 936 & 1.1 & French \\
EmoV-DB~\cite{adigweEmotionalVoicesDatabase2018} & 2018 & Scripted & Expr & 5 & 4 & 6,887 & 9.5 & English \\
JL-Corpus~\cite{jamesOpenSourceEmotional2018} & 2018 & Scripted & Expr & 10 & 4 & 2,400 & 1.4 & English \\
RAVDESS~\cite{livingstoneRyersonAudioVisualDatabase2018} & 2018 & Scripted & Expr & 8 & 24 & 1,440 & 1.5 & English \\
CREMA-D~\cite{caoCREMADCrowdSourcedEmotional2014} & 2014 & Scripted & Expr+Perc & 6 & 91 & 7,442 & 5.3 & English \\
EMOVO~\cite{costantiniEMOVOCorpusItalian2014} & 2014 & Scripted & Expr & 7 & 6 & 588 & 0.5 & Italian \\
SAVEE~\cite{jackson2014savee} & 2014 & Scripted & Expr & 7 & 4 & 480 & 0.5 & English \\
PAVOQUE~\cite{steiner2013pavoque} & 2013 & Scripted & Expr & 5 & 1 & 7,334 & 12.2 & German \\
TESS~\cite{dupuis2010tess} & 2010 & Scripted & Expr & 7 & 2 & 2,800 & 1.6 & English \\
CASIA~\cite{jianhua2008casia} & 2008 & Scripted & Expr & 6 & 6 & 1,200 & 0.6 & Mandarin \\
IEMOCAP~\cite{carlosbussoExpressionPerceptionEmotions2008} & 2008 & Script/Spont & Expr+Perc & 4 & 10 & 5,531 & 7.0 & English \\
eNTERFACE~\cite{martinENTERFACE05AudioVisual2006} & 2006 & Spontaneous & Expr & 6 & 44 & 1,263 & 1.1 & English \\
\bottomrule
\end{tabular*}
\label{tab:dataset_overview}
\endgroup
\vspace{-6mm}
\end{table*}


\section{VoxEmo Benchmark Design}
To fulfill the requirements outlined above, the VoxEmo curates 35 corpora with explicit metadata and standardised partitioning.
\subsection{Data Overview}
Table~\ref{tab:dataset_overview} gives a summary of the datasets used in VoxEmo and presents key dataset attributes. 
As shown 35 datasets released between 2006 and 2025 are included, covering 15 languages. 
Datasets are grouped into in-the-wild (7) and acted (28) recordings and sorted by year of release. 
The table includes year of release, language, number of emotion categories, number of speakers and utterances, hours of speech, and \textbf{Audio source} and \textbf{Label source}. \textbf{Audio source} denotes whether the recording is  ``in-the-wild'', or the scripted/spontaneous style for acted datasets. 
\textbf{Label source} categorises whether annotations target perceived emotion, expressed emotion, or both; as was found in the wild datasets, primarily provide perceived labels, whereas acted datasets typically provide expressed labels, with only a few providing both expressed and perceived labels.

\vspace{-1mm}
\subsection{Soft-Label Ground Truth}
\vspace{-1mm}
Among the datasets annotated by multiple listeners, inter-rater disagreement often reflects meaningful perceptual variance rather than noise \cite{Cowen_2021}. 
To retain this information, the ground truth is modeled as an unsmoothed count-based \textbf{soft-label distribution} $\mathbf{y}$, where denotes the probability simplex over $C$ emotion categories. 
Formally, given a set of $N$ annotations for a specific sample, the target distribution is defined as:
\begin{equation}
    \mathbf{y} = \left[ \frac{n_1}{N}, \frac{n_2}{N}, \dots, \frac{n_C}{N} \right],
\end{equation}
where $n_c$ represents the count of votes for category $c$.  
For instance, in a 4-class scenario $\{ \text{Neutral, Angry, Sad, Happy} \}$ with $N=5$ annotators distributed as $\{1, 2, 2, 0\}$, the resulting label vector is $\mathbf{y} = [0.2, 0.4, 0.4, 0.0]$. 
Since annotator-level ratings are not available across all public emotion databases, soft-label analysis is restricted to the five datasets that include such metadata: CREMA-D, IEMOCAP, MSP-Podcast, BIIC-Podcast, and EmotionTalk.

\begin{table}[!t]
\centering
\fontsize{7}{9}\selectfont
\caption{\small Comparison of the two evaluated speech LLMs.}
\vspace{-3mm}
\begin{tabular}{lcc}
\toprule
 & \textbf{Qwen2-Audio} & \textbf{Audio Flamingo 3} \\
\midrule
Audio encoder & Whisper-large-v3 & Whisper-large-v3 \\
LLM backbone & Qwen-7B & Qwen2.5-7B \\
Training data & 370k hrs, undisclosed & $>$9M QA pairs, released \\
IEMOCAP WA (\%) & 59.2 & 63.8 \\
\bottomrule
\end{tabular}
\label{tab:model_comparison}
\vspace{-6mm}
\end{table}

\subsection{In-Domain Partitioning}
Following the EmoBox protocol \cite{maEmoBoxMultilingualMulticorpus2024}, which provides reproducible, speaker-independent splits for a large number of public SER corpora, datasets with provider-defined splits (e.g., IEMOCAP, MELD) use them directly. 
For the remainder, the partitioning strategy is determined by the number of speakers and the balance of emotion labels across speakers.
Datasets with fewer than 4 speakers or with highly unbalanced emotion-label distributions across speakers are split using stratified random sampling (75\% train, 25\% test; one fold), stratified by emotion category. For datasets with 4--6 speakers, leave-one-speaker-out cross-validation is used, with as many folds as there are distinct speakers. 
For datasets with more than six speakers, speakers are partitioned into four-folds.
All speaker-based schemes ensure that no speaker overlaps between the training and test sets. 
For datasets lacking a pre-defined/official validation set, 20\% of the training partition is held out via stratified sampling.

\vspace{-1mm}
\subsection{Cross-Domain Configurations}
\vspace{-1mm}
To assess cross-corpus generalisation across English-language data, a cross-scenario setting is defined in which label inventories, recording conditions, and \textbf{Label source} (expressed vs. perceived) vary simultaneously.
Unlike traditional classifiers, generative SER facilitates evaluating across these variations by allowing the prompt-specified label set at inference to differ from the one used during training or fine-tuning. 
Four corpora that share a common language but differ systematically in label-set size (4--8 classes), audio source, and label source are evaluated in all source--target combinations: CREMA-D (6-class, acted, expressed labels), IEMOCAP (4-class, acted dialogue, perceived), MELD (7-class, broadcast, perceived), and MSP-Podcast~v2.0 (8-class, podcast, perceived).
The released toolkit also includes cross-lingual configurations; these are not reported here due to space constraints.






\begin{table}[t]
\centering
\fontsize{7}{9}\selectfont 
\setlength{\tabcolsep}{6pt}
\caption{\small Zero-shot prompt template used in VoxEmo. Prompt families are formed by adding instruction blocks to the shared template. For hard-label evaluation, \texttt{FINAL\_LABEL} is parsed as the prediction.}
\vspace{-3mm}
\begin{tabularx}{\columnwidth}{>{\raggedright\arraybackslash}p{0.30\columnwidth}|>{\raggedright\arraybackslash}X}
\hline
\multicolumn{2}{c}{\textbf{Zero-shot Prompt Template}} \\
\hline
\textbf{Task description} &
You are a speech emotion recognition system. You are given an audio clip. \\
\hline
\textbf{Emotion categories} &
\begin{minipage}[t]{\linewidth}\raggedright
Hard-label prompts: Choose exactly one emotion label from this closed set: \texttt{\{labels\}}.\par
Distribution prompt: Consider the following emotion labels: \texttt{\{labels\}}.
\end{minipage} \\
\hline
\multicolumn{2}{l}{\textit{Instruction blocks}} \\
\hline
\textbf{+ Direct decision} &
Decide the emotion directly from the audio. \\
\hline
\textbf{+ Distribution} &
Estimate how likely each emotion is based on the audio. Assign a probability (between 0.0 and 1.0) to every label. The probabilities must sum to 1.0. Also state which single label is most likely. \\
\hline
\textbf{+ ASR transcript} &
Transcribe the spoken content from the audio as accurately as possible. \\
\hline
\textbf{+ Acoustic caption} &
Describe acoustic and paralinguistic cues you hear (e.g., pitch, loudness, speaking rate, pauses, voice quality, non-speech events). \\
\hline
\textbf{+ Reasoning} &
Explain how the transcript content and the acoustic cues together support your final choice. \\
\hline
\textbf{Response Format} &
\begin{minipage}[t]{\linewidth}\raggedright
\setlength{\parskip}{0pt}
\begin{tabular}[t]{@{}l@{}}
\texttt{[ASR\_\allowbreak TRANSCRIPT: <transcript>]}\\[-0.2em]
\texttt{[ACOUSTIC\_\allowbreak CAPTION:]}\\[-0.2em]
\texttt{<description>]}\\[-0.2em]
\texttt{[REASONING: <reasoning>]}\\[-0.2em]
\texttt{[EMOTION\_\allowbreak DISTRIBUTION:}\\[-0.2em]
\texttt{\{"Label1": prob1,}\\[-0.2em]
\texttt{"Label2": prob2, ...\}]}\\[-0.15em]
\texttt{FINAL\_\allowbreak LABEL: <one label from}\\[-0.2em]
\texttt{\{labels\}>}
\end{tabular}
\end{minipage} \\
\hline
\end{tabularx}
\label{tab:zs_prompt_template}
\vspace{-3mm}
\end{table}

\section{Experimental Setup}
\vspace{-1mm}
\subsection{Selection of Speech LLMs}
\label{sec:slm}
\vspace{-1mm}
To maximise reproducibility while minimising implementation-driven variance, the benchmark focuses on open-weight speech LLMs that accept audio inputs and generate text responses under instruction prompts. 
Two representative $\sim$7B audio-language models are selected: Qwen2-Audio-7B-Instruct (Q2A) \cite{chuQwen2AudioTechnicalReport2024} and Audio Flamingo~3 (AF3) \cite{goelAudioFlamingo32025}. Table~\ref{tab:model_comparison} summarises their key attributes.

Q2A's training data composition is not publicly disclosed, making potential data contamination difficult to assess. AF3's released training data explicitly includes SER-relevant corpora such as IEMOCAP and MELD, confirming overlap with benchmark datasets. 
The IEMOCAP WA scores in Table~\ref{tab:model_comparison} are reported by \cite{goelAudioFlamingo32025}. 
Both models are supported by Hugging Face Transformers\footnote{\url{https://github.com/huggingface/transformers}} and PEFT\footnote{\url{https://github.com/huggingface/peft}}, enabling inference and fine-tuning under a single, consistent environment.

\vspace{-1mm}
\subsection{Zero-shot Protocol}
\label{sec:zs-protocal}
\vspace{-1mm}
Zero-shot evaluation follows a fixed inference-time protocol: generation uses greedy decoding, and output parsing rules are shared across all runs; only the prompt template varies.
To assess how auxiliary reasoning modes affect SER performance, five prompt variants are formed by progressively composing instruction blocks onto a shared template (Table~\ref{tab:zs_prompt_template}): ASR transcript (\textbf{T}), acoustic caption (\textbf{A}), and reasoning (\textbf{R}). 
For soft-label evaluation on the five multi-annotator corpora (Section~2.2), each variant additionally includes the \textbf{Distribution} block, which requests a probability distribution over emotion categories.

For hard-label evaluation, the predicted label is extracted from the \texttt{FINAL\_LABEL} field and mapped to the dataset-specific closed label set via case-insensitive matching. 
Samples whose output cannot be mapped are retained as incorrect predictions. 
In soft-label evaluation, the \texttt{EMOTION\_DISTRIBUTION} field is parsed as a JSON object, and mapped to the target label set via alias normalisation, and then re-normalised for probabilities to sum to one. When the parsing process fails, a uniform distribution across all labels is assumed.

Since single-prompt evaluation is sensitive to prompt wording and parsing failures, a vote-based ensemble is adopted that aggregates the 5 hard-label predictions into a single distribution. 
This follows recent findings that various prompt-elicited predictions better reflect human annotation distributions \cite{zhang2026scaling}. 
For sample $k$, let $n_{k,c}$ be the number of prompts predicting class $c$ and $f_k$ the number of parse failures; each failure contributes a uniform vote of $1/C$. The predicted distribution is:
\begin{equation}
    \hat{P}(k,c) = \frac{n_{k,c} + f_k / C}{N_{\text{prompts}}}, \quad N_{\text{prompts}}=5.
\end{equation}

\subsection{Supervised Fine-Tuning}\label{sec:sft-setting}
\vspace{-1mm}
To provide a supervised contrast reference to zero-shot evaluation, both models presented in \ref{sec:slm} are adapted (fine-tuned) to the training sets. 
Following recent practice in SpeechLLM-based SER \cite{su2025reasoning}, LoRA \cite{huLoRALowRankAdaptation2021} adapters  ($r=8$, $\alpha=16$, dropout $0.05$) is applied to the attention projection matrices (q/k/v/o) of the language decoders. 
The audio encoder and the remaining parameters remain frozen, meaning fewer than 0.15\% of the total parameters are updated.  
Training uses AdamW with a learning rate of $1\times10^{-5}$, 10\% warmup, bf16 precision, an effective batch size of 16, for 10 epochs. 
Compared to the multi-block zero-shot template, supervised fine-tuning (SFT) uses a single closed-set instruction prompt that injects the dataset-specific label set \texttt{\{labels\}}; the training target is formatted as \texttt{FINAL\_LABEL: <label>}. 
Inference uses greedy decoding, consistent with the zero-shot setting. Predictions are scored with the same evaluator and metrics as zero-shot runs.
\vspace{-1mm}
\subsection{Evaluation Metrics}
\vspace{-1mm}
For the hard-label case (only one emotion per utterance), the reported standard metrics are Weighted Accuracy (WA), Unweighted Accuracy (UA), Micro-F1, and Macro-F1, computed per dataset over its closed-label set. 
However, these metrics do not reflect annotation disagreement or the natural ambiguity of emotions (more than one emotion per utterance) present in multi-annotator corpora. 
Following \cite{chouMinorityViewsMatter2025}, we additionally report soft-label metrics using a dual-metric approach:
\begin{itemize}
    \item \textbf{Hard-decision Assessment:} The predicted probability distributions are thresholded to compute categorical classification performance, reporting Macro-F1, Micro-F1, and Top-1 Accuracy.
    \item \textbf{Distribution-aware Assessment:} Probabilistic divergence is measured using Kullback-Leibler Divergence (KLD), Jensen-Shannon Divergence (JSD), and Total Variation Distance (TVD), Cosine Similarity (Sim), and Mean Squared Error (MSE) are reported for directional alignment and reconstruction error, respectively.
\end{itemize}

\vspace{-1mm}
\section{Results and Analyses}
\vspace{-1mm}
\subsection{Zero-Shot Evaluation}
\vspace{-1mm}

\begin{table*}[htbp]
\centering
\vspace{-11mm}
\fontsize{7}{9}\selectfont
\caption{VoxEmo zero-shot + SFT results. Q2A\,=\,Qwen2-Audio, AF3\,=\,Audio Flamingo\,3. Prompt abbreviations: T\,=\,ASR transcript, A\,=\,acoustic caption, R\,=\,reasoning. \textbf{\textcolor{red!85}{Bold red}}: best Q2A prompt per dataset; \textbf{\textcolor{red!85}{bold red}}: best AF3 prompt per dataset. \textbf{\textcolor{red!85}{Bold red}}: best score among Q2A SFT / AF3 SFT / EmoBox (by Macro-F1) per dataset. Part 1/2.}
\vspace{-3mm}
\begin{tabular}{l|ccc|ccc|ccc|ccc}
\hline
\textbf{Setting} & \textbf{UA(\%)} & \textbf{WA(\%)} & \textbf{F1(\%)} & \textbf{UA(\%)} & \textbf{WA(\%)} & \textbf{F1(\%)} & \textbf{UA(\%)} & \textbf{WA(\%)} & \textbf{F1(\%)} & \textbf{UA(\%)} & \textbf{WA(\%)} & \textbf{F1(\%)} \\
\cline{2-13}
 & \multicolumn{3}{c|}{\textbf{AESDD (el, 5)}} & \multicolumn{3}{c|}{\textbf{ASED (am, 5)}} & \multicolumn{3}{c|}{\textbf{ASVP-ESD (en/zh, 14)}} & \multicolumn{3}{c}{\textbf{BIIC-Podcast (zh-tw, 8)}} \\
\hline
Q2A Direct & 28.57 & 28.19 & 20.06 & 35.93 & 35.39 & 32.74 & 11.28 & 16.61 & 9.91 & 27.73 & 29.85 & 18.34 \\
Q2A +T & 28.18 & 28.19 & 27.08 & 24.59 & 25 & 18.34 & 7.84 & 11.27 & 10.38 & \textbf{\textcolor{red!85}{28.66}} & \textbf{\textcolor{red!85}{33.7}} & \textbf{\textcolor{red!85}{19.96}} \\
Q2A +A & \textbf{\textcolor{red!85}{39.52}} & \textbf{\textcolor{red!85}{39.6}} & \textbf{\textcolor{red!85}{38.63}} & \textbf{\textcolor{red!85}{41.41}} & \textbf{\textcolor{red!85}{42.21}} & \textbf{\textcolor{red!85}{39.02}} & \textbf{\textcolor{red!85}{20.86}} & \textbf{\textcolor{red!85}{24.55}} & \textbf{\textcolor{red!85}{18.4}} & 21.59 & 39.35 & 16.55 \\
Q2A +T+A & 27.38 & 27.52 & 24.67 & 29.11 & 29.87 & 25.16 & 9.45 & 12.47 & 10.17 & 20.51 & 36.98 & 15.29 \\
Q2A +T+A+R & 26.92 & 26.85 & 25.77 & 30.5 & 31.17 & 26.81 & 14.94 & 21.08 & 15.47 & 24.3 & 40.03 & 18.08 \\
\hline
AF3 Direct & 29.45 & 29.53 & 24.95 & \textbf{\textcolor{red!85}{37.18}} & \textbf{\textcolor{red!85}{38.15}} & \textbf{\textcolor{red!85}{33.28}} & \textbf{\textcolor{red!85}{28.79}} & \textbf{\textcolor{red!85}{34.41}} & \textbf{\textcolor{red!85}{28.36}} & \textbf{\textcolor{red!85}{18.26}} & \textbf{\textcolor{red!85}{41.22}} & \textbf{\textcolor{red!85}{15.43}} \\
AF3 +T & \textbf{\textcolor{red!85}{29.47}} & \textbf{\textcolor{red!85}{29.53}} & \textbf{\textcolor{red!85}{26.65}} & 24.87 & 25.65 & 17.72 & 15.55 & 19.44 & 16.84 & 16.05 & 37.87 & 13.41 \\
AF3 +A & 26.05 & 26.17 & 20.98 & 31.69 & 32.63 & 25.39 & 28.18 & 36.42 & 28.03 & 20.2 & 36.15 & 15.19 \\
AF3 +T+A & 18.05 & 18.12 & 11.97 & 25.71 & 26.62 & 18.25 & 18.65 & 24.75 & 19.27 & 15.07 & 39.69 & 12.58 \\
AF3 +T+A+R & 24.09 & 24.16 & 21.98 & 23.48 & 23.86 & 18.59 & 17.53 & 22.97 & 19.38 & 13.54 & 35.81 & 10.67 \\
\hline
Q2A SFT & 44.3 & 44.3 & 40.09 & 83.99 & 84.09 & 83.96 & 59.03 & 70.38 & 59.51 & \textbf{\textcolor{red!85}{39.5}} & \textbf{\textcolor{red!85}{57.9}} & \textbf{\textcolor{red!85}{40.96}} \\
AF3 SFT & 35.43 & 35.57 & 29.72 & 67.77 & 67.69 & 67.34 & 45.76 & 58.02 & 47.7 & 22.79 & 48.13 & 24.25 \\
EmoBox\cite{maEmoBoxMultilingualMulticorpus2024} & \textbf{\textcolor{red!85}{84.4}} & \textbf{\textcolor{red!85}{84.49}} & \textbf{\textcolor{red!85}{84.19}} & \textbf{\textcolor{red!85}{96.75}} & \textbf{\textcolor{red!85}{96.73}} & \textbf{\textcolor{red!85}{96.74}} & \textbf{\textcolor{red!85}{61.14}} & \textbf{\textcolor{red!85}{71.52}} & \textbf{\textcolor{red!85}{62.08}} & -- & -- & -- \\
\hline
\textbf{Setting} & \multicolumn{3}{c|}{\textbf{CaFE (fr, 7)}} & \multicolumn{3}{c|}{\textbf{CASIA (zh, 6)}} & \multicolumn{3}{c|}{\textbf{CREMA-D (en, 6)}} & \multicolumn{3}{c}{\textbf{EMNS (en, 8)}} \\
\hline
Q2A Direct & 37.1 & 38.57 & 34.12 & 31.25 & 31.25 & 25.31 & \textbf{\textcolor{red!85}{65.94}} & \textbf{\textcolor{red!85}{65.3}} & \textbf{\textcolor{red!85}{66.22}} & 14.02 & 13.4 & 7.21 \\
Q2A +T & 28.67 & 27.24 & 25.22 & 20.08 & 20.08 & 14.49 & 25.71 & 23.87 & 19.5 & \textbf{\textcolor{red!85}{12.46}} & \textbf{\textcolor{red!85}{12.72}} & \textbf{\textcolor{red!85}{7.42}} \\
Q2A +A & \textbf{\textcolor{red!85}{40.58}} & \textbf{\textcolor{red!85}{38.14}} & \textbf{\textcolor{red!85}{37.62}} & \textbf{\textcolor{red!85}{40.5}} & \textbf{\textcolor{red!85}{40.5}} & \textbf{\textcolor{red!85}{35.51}} & 63.5 & 62.82 & 62.14 & 12.75 & 13.06 & 6.9 \\
Q2A +T+A & 40.08 & 36 & 35.41 & 26.75 & 26.75 & 20.71 & 32.95 & 31.52 & 32.61 & 13.19 & 13.4 & 5.01 \\
Q2A +T+A+R & 35.32 & 32.16 & 32.27 & 29.08 & 29.08 & 22.43 & 41.79 & 40.36 & 39.43 & 12.15 & 12.37 & 3.98 \\
\hline
AF3 Direct & \textbf{\textcolor{red!85}{37.5}} & \textbf{\textcolor{red!85}{34.62}} & \textbf{\textcolor{red!85}{31.25}} & 29.75 & 29.75 & 25.31 & \textbf{\textcolor{red!85}{58.49}} & \textbf{\textcolor{red!85}{57.49}} & \textbf{\textcolor{red!85}{59.78}} & \textbf{\textcolor{red!85}{22.58}} & \textbf{\textcolor{red!85}{21.99}} & \textbf{\textcolor{red!85}{18.54}} \\
AF3 +T & 29.27 & 24.14 & 24.82 & 21 & 21 & 13.45 & 38.98 & 37.5 & 39.62 & 13.94 & 14.09 & 6.31 \\
AF3 +A & 32.34 & 27.78 & 27.6 & \textbf{\textcolor{red!85}{31.83}} & \textbf{\textcolor{red!85}{31.83}} & \textbf{\textcolor{red!85}{28.49}} & 50.36 & 49.35 & 50.81 & 16.57 & 16.5 & 10.4 \\
AF3 +T+A & 29.07 & 23.72 & 24.11 & 24.42 & 24.42 & 18.33 & 37.15 & 35.67 & 37.14 & 11.62 & 11.68 & 5.27 \\
AF3 +T+A+R & 27.28 & 22.44 & 20.91 & 23.42 & 23.42 & 17.42 & 28.98 & 27.59 & 28.78 & 9.2 & 9.28 & 4.08 \\
\hline
Q2A SFT & 65.58 & 65.06 & 64.66 & \textbf{\textcolor{red!85}{60}} & \textbf{\textcolor{red!85}{60}} & \textbf{\textcolor{red!85}{56.57}} & \textbf{\textcolor{red!85}{91.17}} & \textbf{\textcolor{red!85}{91.06}} & \textbf{\textcolor{red!85}{91.11}} & 14.41 & 14.43 & 11.08 \\
AF3 SFT & 38.89 & 38.14 & 37.17 & 46.33 & 46.33 & 44.44 & 74.75 & 74.84 & 75.2 & 21.46 & 21.31 & 22.64 \\
EmoBox\cite{maEmoBoxMultilingualMulticorpus2024} & \textbf{\textcolor{red!85}{69.43}} & \textbf{\textcolor{red!85}{68.84}} & \textbf{\textcolor{red!85}{68.06}} & 59.58 & 59.58 & 56.27 & 76.75 & 76.48 & 76.6 & \textbf{\textcolor{red!85}{83.97}} & \textbf{\textcolor{red!85}{84.12}} & \textbf{\textcolor{red!85}{83.97}} \\
\hline
\textbf{Setting} & \multicolumn{3}{c|}{\textbf{EmoDB (de, 7)}} & \multicolumn{3}{c|}{\textbf{EmotionTalk (zh, 7)}} & \multicolumn{3}{c|}{\textbf{EmoV-DB (en, 5)}} & \multicolumn{3}{c}{\textbf{EMOVO (it, 7)}} \\
\hline
Q2A Direct & 69.01 & 67.97 & 64.57 & 27.87 & 17.32 & 19.55 & \textbf{\textcolor{red!85}{68.04}} & \textbf{\textcolor{red!85}{64.05}} & \textbf{\textcolor{red!85}{62.1}} & 23.64 & 23.64 & 19.66 \\
Q2A +T & 39.14 & 33.77 & 36.68 & 35.87 & 38.21 & 27.22 & 40.67 & 39.03 & 37.13 & 16.33 & 16.33 & 12.77 \\
Q2A +A & \textbf{\textcolor{red!85}{75.94}} & \textbf{\textcolor{red!85}{68.4}} & \textbf{\textcolor{red!85}{70.87}} & 29.08 & 37.79 & 22.44 & 57.53 & 58.17 & 55.71 & \textbf{\textcolor{red!85}{27.04}} & \textbf{\textcolor{red!85}{27.04}} & \textbf{\textcolor{red!85}{21.18}} \\
Q2A +T+A & 43.89 & 40.69 & 42.53 & 28.17 & 45.77 & 26.27 & 33.79 & 33.86 & 30.76 & 22.28 & 22.28 & 16.67 \\
Q2A +T+A+R & 47.33 & 42.86 & 45.71 & \textbf{\textcolor{red!85}{30.41}} & \textbf{\textcolor{red!85}{47.07}} & \textbf{\textcolor{red!85}{28.02}} & 42.42 & 41.24 & 38.23 & 23.81 & 23.81 & 18.55 \\
\hline
AF3 Direct & \textbf{\textcolor{red!85}{54.75}} & \textbf{\textcolor{red!85}{54.11}} & \textbf{\textcolor{red!85}{53.71}} & \textbf{\textcolor{red!85}{28.13}} & \textbf{\textcolor{red!85}{45.31}} & \textbf{\textcolor{red!85}{27.64}} & \textbf{\textcolor{red!85}{74.5}} & \textbf{\textcolor{red!85}{75.97}} & \textbf{\textcolor{red!85}{77.49}} & 25.17 & 25.17 & 16.35 \\
AF3 +T & 47.73 & 45.45 & 48.61 & 22.18 & 46.5 & 24.53 & 60.83 & 62.13 & 65.7 & 17.69 & 17.69 & 10.13 \\
AF3 +A & 49 & 48.92 & 51.11 & 24.49 & 45.31 & 22.98 & 65.14 & 66.9 & 69.1 & \textbf{\textcolor{red!85}{24.32}} & \textbf{\textcolor{red!85}{24.32}} & \textbf{\textcolor{red!85}{16.9}} \\
AF3 +T+A & 44.4 & 42.42 & 46.94 & 21.38 & 49.35 & 23.04 & 51.02 & 52.82 & 56.19 & 18.2 & 18.2 & 10.74 \\
AF3 +T+A+R & 37.65 & 37.66 & 41.17 & 18.24 & 42.72 & 19.36 & 45.8 & 47.18 & 51.15 & 18.71 & 18.71 & 11.78 \\
\hline
Q2A SFT & 72.59 & 73.59 & 68 & \textbf{\textcolor{red!85}{59.19}} & \textbf{\textcolor{red!85}{71.28}} & \textbf{\textcolor{red!85}{60.22}} & \textbf{\textcolor{red!85}{99.88}} & \textbf{\textcolor{red!85}{99.88}} & \textbf{\textcolor{red!85}{99.88}} & 33.33 & 33.33 & 30.46 \\
AF3 SFT & 33.38 & 32.03 & 30.7 & 38.86 & 58.68 & 41.3 & 89.8 & 90.29 & 90.45 & 16.84 & 16.84 & 11.23 \\
EmoBox\cite{maEmoBoxMultilingualMulticorpus2024} & \textbf{\textcolor{red!85}{92.58}} & \textbf{\textcolor{red!85}{92.67}} & \textbf{\textcolor{red!85}{92.57}} & -- & -- & -- & 99.44 & 99.47 & 99.45 & \textbf{\textcolor{red!85}{57.82}} & \textbf{\textcolor{red!85}{57.82}} & \textbf{\textcolor{red!85}{56.06}} \\
\hline
\textbf{Setting} & \multicolumn{3}{c|}{\textbf{Emozionalmente (it, 7)}} & \multicolumn{3}{c|}{\textbf{eNTERFACE (en, 6)}} & \multicolumn{3}{c|}{\textbf{ESD (en/zh, 5)}} & \multicolumn{3}{c}{\textbf{IEMOCAP (en, 4)}} \\
\hline
Q2A Direct & 23 & 23 & 17.84 & 43.27 & 43.27 & 40.09 & 38.18 & 38.18 & 35.05 & 54.38 & 56.97 & 51.63 \\
Q2A +T & 18.99 & 18.99 & 14.46 & \textbf{\textcolor{red!85}{63.14}} & \textbf{\textcolor{red!85}{63.14}} & \textbf{\textcolor{red!85}{60.26}} & 25.66 & 25.66 & 21.52 & 44.58 & 46.9 & 46 \\
Q2A +A & \textbf{\textcolor{red!85}{27.12}} & \textbf{\textcolor{red!85}{27.12}} & \textbf{\textcolor{red!85}{22.32}} & 43.91 & 43.91 & 38.09 & \textbf{\textcolor{red!85}{42.07}} & \textbf{\textcolor{red!85}{42.07}} & \textbf{\textcolor{red!85}{40.3}} & \textbf{\textcolor{red!85}{57.33}} & \textbf{\textcolor{red!85}{59.09}} & \textbf{\textcolor{red!85}{58.19}} \\
Q2A +T+A & 21.6 & 21.6 & 16.98 & 53.53 & 53.53 & 52.36 & 28.47 & 28.47 & 24.3 & 39.92 & 43.85 & 40.03 \\
Q2A +T+A+R & 23.23 & 23.23 & 18.55 & 56.09 & 56.09 & 55.19 & 33.96 & 33.96 & 30.06 & 41.25 & 44.9 & 41.81 \\
\hline
AF3 Direct & \textbf{\textcolor{red!85}{24.56}} & \textbf{\textcolor{red!85}{24.56}} & \textbf{\textcolor{red!85}{20.72}} & \textbf{\textcolor{red!85}{53.53}} & \textbf{\textcolor{red!85}{53.53}} & \textbf{\textcolor{red!85}{56.12}} & 37.65 & 37.65 & 32.34 & \textbf{\textcolor{red!85}{62.6}} & \textbf{\textcolor{red!85}{63.88}} & \textbf{\textcolor{red!85}{63.97}} \\
AF3 +T & 16.26 & 16.26 & 8.51 & 48.72 & 48.72 & 49.05 & 28.43 & 28.43 & 22.66 & 46.57 & 48.88 & 49.9 \\
AF3 +A & 22.01 & 22.01 & 15.35 & 49.04 & 49.04 & 48.68 & \textbf{\textcolor{red!85}{36.69}} & \textbf{\textcolor{red!85}{36.69}} & \textbf{\textcolor{red!85}{34.15}} & 60.1 & 61.4 & 61.09 \\
AF3 +T+A & 16.9 & 16.9 & 10.22 & 44.55 & 44.55 & 46.11 & 28.31 & 28.31 & 23.28 & 39.08 & 42.07 & 41.11 \\
AF3 +T+A+R & 15.45 & 15.45 & 9.14 & 38.14 & 38.14 & 39.4 & 24.58 & 24.58 & 21.26 & 43.11 & 45.99 & 43.38 \\
\hline
Q2A SFT & 74.62 & 74.62 & 74.61 & 95.83 & 95.83 & 95.85 & 83.7 & 83.7 & 83.58 & \textbf{\textcolor{red!85}{82.65}} & \textbf{\textcolor{red!85}{82.13}} & \textbf{\textcolor{red!85}{82.43}} \\
AF3 SFT & 54.65 & 54.65 & 54.28 & 60.9 & 60.9 & 61.19 & 71.72 & 71.72 & 71.75 & 63.46 & 65.01 & 65.15 \\
EmoBox\cite{maEmoBoxMultilingualMulticorpus2024}& \textbf{\textcolor{red!85}{76.91}} & \textbf{\textcolor{red!85}{76.91}} & \textbf{\textcolor{red!85}{76.9}} & \textbf{\textcolor{red!85}{97.69}} & \textbf{\textcolor{red!85}{97.68}} & \textbf{\textcolor{red!85}{97.68}} & \textbf{\textcolor{red!85}{84.62}} & \textbf{\textcolor{red!85}{84.62}} & \textbf{\textcolor{red!85}{84.33}} & 73.54 & 72.86 & 73.11 \\
\hline
\textbf{Setting} & \multicolumn{3}{c|}{\textbf{JL-Corpus (en, 10)}} & \multicolumn{3}{c|}{\textbf{M3ED (zh, 7)}} & \multicolumn{3}{c|}{\textbf{MEAD (en, 8)}} & \multicolumn{3}{c}{\textbf{MELD (en, 7)}} \\
\hline
Q2A Direct & \textbf{\textcolor{red!85}{20.79}} & \textbf{\textcolor{red!85}{20.79}} & \textbf{\textcolor{red!85}{14.77}} & 17.18 & 20.13 & 11.38 & \textbf{\textcolor{red!85}{16.14}} & \textbf{\textcolor{red!85}{16.76}} & \textbf{\textcolor{red!85}{13.38}} & 31.68 & 42.91 & 25.97 \\
Q2A +T & 12.88 & 12.88 & 9.52 & \textbf{\textcolor{red!85}{21.49}} & \textbf{\textcolor{red!85}{36.95}} & \textbf{\textcolor{red!85}{18.93}} & 16.05 & 14.24 & 13.22 & \textbf{\textcolor{red!85}{32.23}} & \textbf{\textcolor{red!85}{42.98}} & \textbf{\textcolor{red!85}{28.06}} \\
Q2A +A & 16.46 & 16.46 & 9.68 & 18.63 & 27.51 & 12.81 & 15.1 & 11.04 & 8.88 & 29.07 & 47.39 & 24.81 \\
Q2A +T+A & 11.08 & 11.08 & 4.19 & 15.39 & 37.99 & 12.97 & 14.71 & 9.88 & 8.65 & 27.37 & 46.24 & 25.49 \\
Q2A +T+A+R & 12.96 & 12.96 & 8.28 & 17.93 & 40.71 & 15.2 & 14.06 & 9.41 & 8.28 & 27.75 & 48.04 & 26.48 \\
\hline
AF3 Direct & \textbf{\textcolor{red!85}{46.5}} & \textbf{\textcolor{red!85}{46.5}} & \textbf{\textcolor{red!85}{49.06}} & 18.25 & 33.85 & 14.88 & \textbf{\textcolor{red!85}{15.81}} & \textbf{\textcolor{red!85}{11.17}} & \textbf{\textcolor{red!85}{8.92}} & \textbf{\textcolor{red!85}{29.77}} & \textbf{\textcolor{red!85}{44.75}} & \textbf{\textcolor{red!85}{29.83}} \\
AF3 +T & 22.17 & 22.17 & 21.82 & 14.54 & 40.28 & 12.38 & 13.79 & 8.06 & 5.48 & 25.53 & 49.16 & 27.52 \\
AF3 +A & 32.58 & 32.58 & 33.25 & \textbf{\textcolor{red!85}{19.23}} & \textbf{\textcolor{red!85}{39.47}} & \textbf{\textcolor{red!85}{15.04}} & 14.87 & 9.37 & 7.56 & 27.82 & 47.85 & 27.91 \\
AF3 +T+A & 21.67 & 21.67 & 21.48 & 16.05 & 42.83 & 13.04 & 12.66 & 6.85 & 3.83 & 20.61 & 49.46 & 21.63 \\
AF3 +T+A+R & 20.88 & 20.88 & 20.28 & 15.23 & 40.26 & 12.91 & 12.58 & 6.96 & 4.3 & 22.74 & 50.15 & 23.39 \\
\hline
Q2A SFT & 44.33 & 44.33 & 41.83 & \textbf{\textcolor{red!85}{37.57}} & \textbf{\textcolor{red!85}{54.95}} & \textbf{\textcolor{red!85}{38.24}} & 69.36 & 70.26 & 69.63 & \textbf{\textcolor{red!85}{37.85}} & \textbf{\textcolor{red!85}{56.33}} & \textbf{\textcolor{red!85}{39.26}} \\
AF3 SFT & 37.08 & 37.08 & 37.09 & 25.91 & 50.24 & 26.78 & 48.39 & 49.36 & 48.94 & 24.12 & 48.54 & 25.74 \\
EmoBox\cite{maEmoBoxMultilingualMulticorpus2024} & \textbf{\textcolor{red!85}{66.71}} & \textbf{\textcolor{red!85}{66.71}} & \textbf{\textcolor{red!85}{65.19}} & 32.84 & 49.42 & 33.76 & \textbf{\textcolor{red!85}{81.27}} & \textbf{\textcolor{red!85}{82.03}} & \textbf{\textcolor{red!85}{81.43}} & 31.54 & 51.89 & 32.95 \\
\hline
\textbf{Setting} & \multicolumn{3}{c|}{\textbf{MESD (es, 6)}} & \multicolumn{3}{c|}{\textbf{MSP-Podcast(test1) (en, 8)}} & \multicolumn{3}{c|}{\textbf{MSP-Podcast(test2) (en, 8)}} & \multicolumn{3}{c}{\textbf{nEMO (pl, 6)}} \\
\hline
Q2A Direct & 23.27 & 23.36 & 17.75 & 22.23 & 17.2 & 12.28 & 17.99 & 16.53 & 9.18 & 21.88 & 22.43 & 14.3 \\
Q2A +T & 22.62 & 22.43 & 19.74 & \textbf{\textcolor{red!85}{24.84}} & \textbf{\textcolor{red!85}{31.42}} & \textbf{\textcolor{red!85}{20.35}} & \textbf{\textcolor{red!85}{19.6}} & \textbf{\textcolor{red!85}{32.18}} & \textbf{\textcolor{red!85}{15.35}} & 15.38 & 16.18 & 11.35 \\
Q2A +A & \textbf{\textcolor{red!85}{21.45}} & \textbf{\textcolor{red!85}{21.5}} & \textbf{\textcolor{red!85}{20.54}} & 21.79 & 38.1 & 16.89 & 16.86 & 55.48 & 14.69 & \textbf{\textcolor{red!85}{21.34}} & \textbf{\textcolor{red!85}{22.07}} & \textbf{\textcolor{red!85}{16.89}} \\
Q2A +T+A & 20.74 & 20.56 & 17.65 & 20.91 & 35.59 & 17.72 & 15.99 & 43.15 & 13.32 & 17.22 & 18.14 & 12.36 \\
Q2A +T+A+R & 18.85 & 18.69 & 15.56 & 20.62 & 37.19 & 18.78 & 16.05 & 45.54 & 14.13 & 17.66 & 18.59 & 13.17 \\
\hline
AF3 Direct & \textbf{\textcolor{red!85}{26.5}} & \textbf{\textcolor{red!85}{26.17}} & \textbf{\textcolor{red!85}{22.32}} & \textbf{\textcolor{red!85}{29.11}} & \textbf{\textcolor{red!85}{47.66}} & \textbf{\textcolor{red!85}{28.67}} & \textbf{\textcolor{red!85}{20.58}} & \textbf{\textcolor{red!85}{51.42}} & \textbf{\textcolor{red!85}{18.78}} & \textbf{\textcolor{red!85}{17.88}} & \textbf{\textcolor{red!85}{19.12}} & \textbf{\textcolor{red!85}{9.04}} \\
AF3 +T & 16.59 & 16.36 & 12.69 & 21.35 & 38.34 & 18.77 & 15.7 & 51.26 & 13.3 & 15.45 & 16.71 & 5.5 \\
AF3 +A & 21.81 & 21.5 & 15.1 & 23.57 & 42.34 & 21.38 & 18.4 & 56.36 & 16.57 & 17.57 & 18.86 & 7.75 \\
AF3 +T+A & 17.12 & 16.82 & 6.84 & 16.87 & 35.99 & 14.81 & 14.4 & 52.78 & 12.2 & 16.03 & 17.34 & 5.71 \\
AF3 +T+A+R & 17.59 & 17.29 & 7.65 & 16.29 & 34.06 & 14.33 & 13.05 & 47.07 & 10.72 & 15.54 & 16.8 & 5.67 \\
\hline
Q2A SFT & 34.29 & 34.11 & 34.3 & \textbf{\textcolor{red!85}{32.08}} & \textbf{\textcolor{red!85}{45.3}} & \textbf{\textcolor{red!85}{29.18}} & \textbf{\textcolor{red!85}{23.24}} & \textbf{\textcolor{red!85}{48.26}} & \textbf{\textcolor{red!85}{18.21}} & \textbf{\textcolor{red!85}{78.16}} & \textbf{\textcolor{red!85}{78.28}} & \textbf{\textcolor{red!85}{78.07}} \\
AF3 SFT & 36.2 & 35.98 & 31.62 & 22.01 & 38.54 & 19.08 & 18.75 & 48.89 & 14.86 & 61.86 & 62.2 & 61.56 \\
EmoBox\cite{maEmoBoxMultilingualMulticorpus2024} & \textbf{\textcolor{red!85}{69.78}} & \textbf{\textcolor{red!85}{69.67}} & \textbf{\textcolor{red!85}{69.64}} & -- & -- & -- & -- & -- & -- & -- & -- & -- \\
\hline
\end{tabular}
\label{tab:emobox2_scoreboard_1}
\end{table*}

\begin{table*}[!t]
\centering
\fontsize{7}{9}\selectfont
\caption{VoxEmo zero-shot + SFT results. Q2A\,=\,Qwen2-Audio, AF3\,=\,Audio Flamingo\,3. Prompt abbreviations: T\,=\,ASR transcript, A\,=\,acoustic caption, R\,=\,reasoning. \textbf{\textcolor{red!85}{Bold red}}: best Q2A prompt per dataset; \textbf{\textcolor{red!85}{bold red}}: best AF3 prompt per dataset. \textbf{\textcolor{red!85}{Bold red}}: best score among Q2A SFT / AF3 SFT / EmoBox\cite{maEmoBoxMultilingualMulticorpus2024} (by MacroF1) per dataset. Part 2/2.}
\vspace{-3mm}
\begin{tabular}{l|ccc|ccc|ccc|ccc}
\hline
\textbf{Setting} & \textbf{UA(\%)} & \textbf{WA(\%)} & \textbf{F1(\%)} & \textbf{UA(\%)} & \textbf{WA(\%)} & \textbf{F1(\%)} & \textbf{UA(\%)} & \textbf{WA(\%)} & \textbf{F1(\%)} & \textbf{UA(\%)} & \textbf{WA(\%)} & \textbf{F1(\%)} \\
\cline{2-13}
 & \multicolumn{3}{c|}{\textbf{Oreau (fr, 7)}} & \multicolumn{3}{c|}{\textbf{PAVOQUE (de, 5)}} & \multicolumn{3}{c|}{\textbf{Polish (pl, 3)}} & \multicolumn{3}{c}{\textbf{RAVDESS (en, 8)}} \\
\hline
Q2A Direct & 46.22 & 47.12 & 39.85 & \textbf{\textcolor{red!85}{47.56}} & \textbf{\textcolor{red!85}{62.77}} & \textbf{\textcolor{red!85}{46.09}} & 34.89 & 34.89 & 21.97 & \textbf{\textcolor{red!85}{72.4}} & \textbf{\textcolor{red!85}{72.43}} & \textbf{\textcolor{red!85}{67.55}} \\
Q2A +T & 39.84 & 41.35 & 37.53 & 27.72 & 53 & 29.07 & 34 & 34 & 27.91 & 49.15 & 45.76 & 44.54 \\
Q2A +A & 33.12 & 35.58 & 25.72 & 45.23 & 67.52 & 45.27 & \textbf{\textcolor{red!85}{34.67}} & \textbf{\textcolor{red!85}{34.67}} & \textbf{\textcolor{red!85}{30.14}} & 68.29 & 66.39 & 62.73 \\
Q2A +T+A & 41.89 & 44.23 & 41.82 & 31.14 & 59.61 & 33.85 & 35.11 & 35.11 & 25.8 & 49.8 & 47.01 & 48.57 \\
Q2A +T+A+R & \textbf{\textcolor{red!85}{45.52}} & \textbf{\textcolor{red!85}{48.08}} & \textbf{\textcolor{red!85}{46.17}} & 34.54 & 60.26 & 36.3 & 34.44 & 34.44 & 27.34 & 45.12 & 44.93 & 44 \\
\hline
AF3 Direct & 33.62 & 35.58 & 35.49 & \textbf{\textcolor{red!85}{38.3}} & \textbf{\textcolor{red!85}{67.3}} & \textbf{\textcolor{red!85}{41.11}} & 34.67 & 34.67 & 19.25 & \textbf{\textcolor{red!85}{47.85}} & \textbf{\textcolor{red!85}{45.42}} & \textbf{\textcolor{red!85}{44.66}} \\
AF3 +T & \textbf{\textcolor{red!85}{36.84}} & \textbf{\textcolor{red!85}{39.42}} & \textbf{\textcolor{red!85}{37.12}} & 25.58 & 55.51 & 27.94 & 33.11 & 33.11 & 19.59 & 34.9 & 30.56 & 32.92 \\
AF3 +A & 30.75 & 32.69 & 30.24 & 31.32 & 64.46 & 32.98 & 34 & 34 & 18.33 & 42.12 & 38.82 & 40.17 \\
AF3 +T+A & 28.17 & 30.77 & 24.81 & 23.46 & 57.97 & 22.97 & 32.44 & 32.44 & 19.45 & 35.03 & 30.9 & 34.96 \\
AF3 +T+A+R & 32.49 & 34.62 & 32.55 & 20 & 46.56 & 22.43 & \textbf{\textcolor{red!85}{36.22}} & \textbf{\textcolor{red!85}{36.22}} & \textbf{\textcolor{red!85}{29.01}} & 33.46 & 30.14 & 34.57 \\
\hline
Q2A SFT & 50.34 & 51.92 & 50.85 & \textbf{\textcolor{red!85}{89.84}} & \textbf{\textcolor{red!85}{94.21}} & \textbf{\textcolor{red!85}{89.95}} & 44.67 & 44.67 & 37.02 & \textbf{\textcolor{red!85}{87.76}} & \textbf{\textcolor{red!85}{90.83}} & \textbf{\textcolor{red!85}{87.81}} \\
AF3 SFT & 43.83 & 46.15 & 46.13 & 72.06 & 84.77 & 76.31 & 36 & 36 & 28.6 & 34.44 & 35.7 & 34.77 \\
EmoBox\cite{maEmoBoxMultilingualMulticorpus2024} & \textbf{\textcolor{red!85}{84.48}} & \textbf{\textcolor{red!85}{84.79}} & \textbf{\textcolor{red!85}{85.01}} & 87.73 & 93.4 & 88.43 & \textbf{\textcolor{red!85}{83.27}} & \textbf{\textcolor{red!85}{83.27}} & \textbf{\textcolor{red!85}{82.77}} & 75.32 & 75.87 & 75.19 \\
\hline
\textbf{Setting} & \multicolumn{3}{c|}{\textbf{RESD (ru, 7)}} & \multicolumn{3}{c|}{\textbf{SAVEE (en, 7)}} & \multicolumn{3}{c|}{\textbf{ShEMO (fa, 6)}} & \multicolumn{3}{c}{\textbf{SUBESCO (bn, 7)}} \\
\hline
Q2A Direct & 22.38 & 22.47 & 16.74 & 54.41 & 56.04 & 51.73 & 47.3 & 49.79 & 43.56 & 30.39 & 30.39 & 24.84 \\
Q2A +T & 18.54 & 18.35 & 17.04 & 28.33 & 31.04 & 27.1 & 26.23 & 30.98 & 24.88 & 17.69 & 17.69 & 15.13 \\
Q2A +A & 23.72 & 22.1 & 19.22 & \textbf{\textcolor{red!85}{58.33}} & \textbf{\textcolor{red!85}{61.88}} & \textbf{\textcolor{red!85}{58.51}} & \textbf{\textcolor{red!85}{69.85}} & \textbf{\textcolor{red!85}{71.57}} & \textbf{\textcolor{red!85}{59.43}} & \textbf{\textcolor{red!85}{37.23}} & \textbf{\textcolor{red!85}{37.23}} & \textbf{\textcolor{red!85}{32.39}} \\
Q2A +T+A & 20.02 & 20.23 & 18.15 & 42.38 & 47.5 & 43.47 & 33.91 & 43.14 & 32.01 & 25.96 & 25.96 & 22.12 \\
Q2A +T+A+R & \textbf{\textcolor{red!85}{23.23}} & \textbf{\textcolor{red!85}{23.22}} & \textbf{\textcolor{red!85}{22.17}} & 45.48 & 50.62 & 45.65 & 41.9 & 46.96 & 36.44 & 27.56 & 27.56 & 24.07 \\
\hline
AF3 Direct & \textbf{\textcolor{red!85}{22.8}} & \textbf{\textcolor{red!85}{23.22}} & \textbf{\textcolor{red!85}{17.14}} & \textbf{\textcolor{red!85}{53.1}} & \textbf{\textcolor{red!85}{55.42}} & \textbf{\textcolor{red!85}{54.57}} & 52.46 & 60.96 & 48.53 & \textbf{\textcolor{red!85}{24.73}} & \textbf{\textcolor{red!85}{24.73}} & \textbf{\textcolor{red!85}{19.18}} \\
AF3 +T & 14.71 & 14.61 & 11.01 & 34.64 & 42.5 & 34.71 & 30.7 & 47.1 & 33.77 & 20.57 & 20.57 & 14.24 \\
AF3 +A & 21.98 & 22.47 & 15.76 & 46.67 & 52.92 & 48.44 & \textbf{\textcolor{red!85}{49.06}} & \textbf{\textcolor{red!85}{64.64}} & \textbf{\textcolor{red!85}{50.31}} & 24.51 & 24.51 & 17.77 \\
AF3 +T+A & 13.89 & 13.86 & 7.87 & 35.12 & 42.29 & 37.06 & 24.4 & 40.03 & 27.12 & 19.93 & 19.93 & 14.12 \\
AF3 +T+A+R & 12.02 & 11.98 & 6.66 & 30.83 & 37.29 & 32.15 & 24.26 & 33.95 & 26.36 & 13.6 & 13.6 & 13.04 \\
\hline
Q2A SFT & 28.88 & 30.71 & 27.18 & 64.64 & 65 & 62.3 & \textbf{\textcolor{red!85}{87.16}} & \textbf{\textcolor{red!85}{95.19}} & \textbf{\textcolor{red!85}{88.28}} & \textbf{\textcolor{red!85}{76.29}} & \textbf{\textcolor{red!85}{76.29}} & \textbf{\textcolor{red!85}{76.16}} \\
AF3 SFT & 20.6 & 20.97 & 19.92 & 33.34 & 37.71 & 34.57 & 57.15 & 75.95 & 60.62 & 31.36 & 31.36 & 31.46 \\
EmoBox\cite{maEmoBoxMultilingualMulticorpus2024} & \textbf{\textcolor{red!85}{55.87}} & \textbf{\textcolor{red!85}{56.47}} & \textbf{\textcolor{red!85}{55.82}} & \textbf{\textcolor{red!85}{75.65}} & \textbf{\textcolor{red!85}{78.25}} & \textbf{\textcolor{red!85}{78.38}} & 80.23 & 89.55 & 82.94 & 73.05 & 73.05 & 72.94 \\
\hline
\textbf{Setting} & \multicolumn{3}{c|}{\textbf{TESS (en, 7)}} & \multicolumn{3}{c|}{\textbf{Thai-ser (th, 5)}} & \multicolumn{3}{c|}{\textbf{TurEV-DB (tr, 4)}} & \multicolumn{3}{c}{\textbf{URDU (ur, 4)}} \\
\hline
Q2A Direct & 71.86 & 71.86 & 73.66 & \textbf{\textcolor{red!85}{29.21}} & \textbf{\textcolor{red!85}{29.17}} & \textbf{\textcolor{red!85}{25.49}} & \textbf{\textcolor{red!85}{28.11}} & \textbf{\textcolor{red!85}{28.93}} & \textbf{\textcolor{red!85}{23.83}} & 38 & 38 & 35.52 \\
Q2A +T & 36.86 & 36.86 & 36.02 & 17.09 & 16.45 & 13.62 & 21.62 & 21.76 & 20.44 & 22 & 22 & 19.05 \\
Q2A +A & \textbf{\textcolor{red!85}{77}} & \textbf{\textcolor{red!85}{77}} & \textbf{\textcolor{red!85}{73.89}} & 25.42 & 24.59 & 21.56 & 26.79 & 27.78 & 21.39 & \textbf{\textcolor{red!85}{36}} & \textbf{\textcolor{red!85}{36}} & \textbf{\textcolor{red!85}{35.97}} \\
Q2A +T+A & 26.86 & 26.86 & 24.92 & 20.12 & 19.09 & 14.03 & 22.77 & 21.76 & 15.14 & 27 & 27 & 21.4 \\
Q2A +T+A+R & 58.71 & 58.71 & 58.44 & 21.12 & 20.09 & 15.18 & 22.45 & 21.53 & 14.14 & 24 & 24 & 19.78 \\
\hline
AF3 Direct & \textbf{\textcolor{red!85}{59.14}} & \textbf{\textcolor{red!85}{59.14}} & \textbf{\textcolor{red!85}{61.36}} & \textbf{\textcolor{red!85}{30.8}} & \textbf{\textcolor{red!85}{29.93}} & \textbf{\textcolor{red!85}{23.54}} & 28.67 & 31.25 & 19.43 & 38 & 38 & 32.2 \\
AF3 +T & 51.86 & 51.86 & 57 & 23.79 & 22.11 & 15.81 & 23.48 & 23.15 & 18.88 & 29 & 29 & 20.93 \\
AF3 +A & 55.57 & 55.57 & 59.58 & 28.35 & 27.78 & 20.39 & \textbf{\textcolor{red!85}{28.36}} & \textbf{\textcolor{red!85}{28.47}} & \textbf{\textcolor{red!85}{22}} & \textbf{\textcolor{red!85}{42}} & \textbf{\textcolor{red!85}{42}} & \textbf{\textcolor{red!85}{36.85}} \\
AF3 +T+A & 53.43 & 53.43 & 58.55 & 22.73 & 21.47 & 14.31 & 22.15 & 21.07 & 11.78 & 34 & 34 & 24.15 \\
AF3 +T+A+R & 47.14 & 47.14 & 52.88 & 19.92 & 18.75 & 12.46 & 22.04 & 21.07 & 12.01 & 36 & 36 & 25 \\
\hline
Q2A SFT & 99.86 & 99.86 & 99.86 & \textbf{\textcolor{red!85}{72.42}} & \textbf{\textcolor{red!85}{72.79}} & \textbf{\textcolor{red!85}{72.3}} & 51.88 & 53.24 & 51.36 & 41 & 41 & 35.46 \\
AF3 SFT & 82.71 & 82.71 & 83.19 & 51.95 & 52.45 & 51.7 & 42.49 & 43.75 & 42.52 & 41 & 41 & 33.75 \\
EmoBox\cite{maEmoBoxMultilingualMulticorpus2024} & \textbf{\textcolor{red!85}{99.96}} & \textbf{\textcolor{red!85}{99.96}} & \textbf{\textcolor{red!85}{99.96}} & -- & -- & -- & \textbf{\textcolor{red!85}{81.32}} & \textbf{\textcolor{red!85}{81.58}} & \textbf{\textcolor{red!85}{81.31}} & \textbf{\textcolor{red!85}{88.41}} & \textbf{\textcolor{red!85}{88.41}} & \textbf{\textcolor{red!85}{88.4}} \\
\hline
\end{tabular}
\label{tab:emobox2_scoreboard_2}
\vspace{-2mm}
\end{table*}

Results in the tables~\ref{tab:emobox2_scoreboard_1}--\ref{tab:emobox2_scoreboard_2} show Macro-F1, UA, and WA results for both Q2A and AF3, under five zero-shot prompt variants on 35~datasets. 
As can be observed prompt choice has a substantial effect on performance in almost all cases. 
For Q2A, the Macro-F1 difference between the best and worst variants exceeds 20 on 11 datasets, reaching a maximum spread of 49 on TESS. 
The optimal variant differs across models and datasets: AF3 performs better with the Direct prompt on 25/35~datasets, whereas Q2A most often benefits from the acoustic-caption variant (+A, 18/35). 
In absolute terms, zero-shot performance remains limited, with the best-prompt Macro-F1 falling below 20\% on eight datasets for both models and staying below the EmoBox supervised baselines on all 30 comparable datasets. 
Under best-prompt comparison, Q2A outperforms AF3 on 26 of the 35~datasets, though this advantage should be interpreted alongside the training-data overlap discussed in Section \ref{sec:slm}. 
The following subsections provide further analysis on how individual instruction blocks contribute to this variation.

\vspace{-1mm}
\subsubsection{Effect of Acoustic Caption}\label{sec:Effect-of-Acoustic-Caption}
\vspace{-1mm}
For Q2A, the acoustic-caption prompt (+A) most frequently yields the highest Macro-F1, for 18 out of 35~datasets. This advantage, however, is concentrated on acted corpora (16/28 acted vs 2/7 in-the-wild). 
For scripted recordings, transcript content is fixed across emotion categories, leaving paralinguistic cues as the primary discriminatory signal.  
Explicitly eliciting an acoustic description appears to help the model attend to these cues. 
Within the acted subset, the benefit is mostly visible for non-English data (13/18 non-EN vs 3/10 EN acted), suggesting that the language-agnostic nature of paralinguistic descriptions additionally compensates for weaker cross-lingual text understanding. 
AF3 is considerably less responsive: The Direct prompt remains its best variant on 25/35~datasets, and +A lowers Macro-F1 relative to Direct for 28/35~datasets. 
The difference between the two models may stem from differences in the composition of the training data or in how acoustic descriptions are integrated with the language decoder.

\vspace{-1mm}
\subsubsection{Effect of ASR Transcripts}
\vspace{-1mm}
Adding the transcript prompt (+T) lowers Macro-F1 on the majority of the acted datasets for both models (Q2A 22/28, AF3 25/28), with Q2A Macro-F1 decreasing by up to 46.7 on CREMA-D and 37.6 on TESS. As discussed in Section \ref{sec:Effect-of-Acoustic-Caption}, scripted recordings fix transcript content across emotion categories, so the added lexical signal is non-discriminative and may introduce noise. 
In in-the-wild corpora, the pattern reverses for Q2A, improving across 6 of 7 datasets, with URDU as the sole exception. 
These gains focus on English and Mandarin, the two languages most represented in Q2A's training data; the URDU result is consistent with the limited coverage of underrepresented languages in current speech LLMs. 
AF3, by contrast, does not benefit from +T on any in-the-wild dataset, with all seven deltas negative, including English corpora such as MELD. 
While AF3's training data is predominantly English, this failure on English corpora points to a model-level limitation in leveraging transcript content for emotion discrimination, rather than a language coverage issue alone. 
The acted--naturalistic split in +T effectiveness aligns directly with the \textbf{Audio source} metadata in Table~\ref{tab:dataset_overview}, confirming that prompt sensitivity is systematically shaped by corpus construction.

\vspace{-1mm}
\subsubsection{Combined Prompts}
\vspace{-1mm}
Combining transcript and acoustic captions in a single prompt (+T+A) does not seem to add value for the Macro-F1 of either model across any of the 35 datasets. 
Across the 70 model--dataset pairs, the optimal variant is a single-block or Direct prompt in 66 cases (Q2A 32/35, AF3 34/35). 
Appending a reasoning step (+T+A+R) partially recovers performance for Q2A, improving over +T+A on 28/35 datasets, but it exceeds the best single-block prompt's performance on only 3 datasets. 
Single-block prompts thus consistently outperform multi-block alternatives within this benchmark, a pattern that may reflect capacity limitations of the two 7\, B-parameter models in integrating multiple information sources, though this may not hold for larger or differently trained speech LLMs.

\vspace{-1mm}
\subsection{Supervised Fine-tuning}
\vspace{-1mm}
The SFT rows in Tables~\ref{tab:emobox2_scoreboard_1} -- \ref{tab:emobox2_scoreboard_2} report the Macro-F1 test for Q2A and AF3 after LoRA fine-tuning (Sec.~\ref{sec:sft-setting}).
SFT improves Q2A over its best zero-shot prompt on 34 of 36 evaluation splits, with a mean absolute Macro-F1 gain of 23.7. 
Gains exceed 40 points on 9~datasets, including cases where zero-shot Macro-F1 was below 25. The improvement is more pronounced on the acted corpora than on the in-the-wild data, yielding a mean SFT Macro-F1 of 66.0 and 40.1, respectively. 
Only EmoDB ($-$2.9) and URDU ($-$0.5) show slight degradation.

On the 30~datasets with a comparable EmoBox supervised reference~\cite{maEmoBoxMultilingualMulticorpus2024}, Q2A surpasses the reference on 10 and matches it within 3~Macro-F1 on a further 5 (e.g., TESS, ESD, eNTERFACE). 
This performance correlates with the size of the dataset: Q2A exceeds the reference on 7 of 11~datasets with at least 5,000~utterances, but on none of the 9~datasets with fewer than 1,000. 
Among in-the-wild corpora, Q2A exceeds the reference on MELD (+6.3) and M3ED (+4.4), two datasets where the EmoBox supervised baseline itself remains below 34~Macro-F1, suggesting that speech LLM fine-tuning can be competitive on naturalistic data where conventional approaches also struggle. 
Nevertheless, 10~evaluation splits remain below 40~Macro-F1 after SFT. The five acted cases are all corpora with fewer than 1,500~utterances (EMNS, RESD, EMOVO, MESD, Polish), where the limited training data likely limits the improvement for LoRA adaptation. 
The five in-the-wild cases span a wider range of corpus sizes, from URDU (400~utterances) to MSP-Podcast (213k), suggesting that factors beyond data volume contribute to difficulty on naturalistic speech.

Under the same LoRA configuration, AF3 improves on 26 of 36~splits but with a mean gain of 10.3, less than half of Q2A's 23.7. The remaining 10~splits degrade, with the largest drops on EmoDB ($-$23.0), SAVEE ($-$20.0), and JL-Corpus ($-$12.0). AF3 does not surpass the EmoBox reference on any of the 30~comparable datasets, with CREMA-D ($-$1.4) as the closest. These results reflect a single LoRA configuration ($r{=}8$); AF3 may benefit from different hyperparameters, a point we return to in Section~5.

Overall, SFT substantially narrows the gap between speech LLMs and traditional supervised baselines, with Q2A reaching parity or better on 15 of 30~comparable datasets. 
The remaining deficit concentrates in small-act corpora and depends heavily on the choice of foundation model, as the widening of Q2A's advantage from 26/35 datasets in zero-shot to 35/36 after SFT illustrates.

\begin{table*}[!t]
\centering
\fontsize{7}{9}\selectfont
\caption{Zero-shot soft-label evaluation results. We compare the single direct prompt (\texttt{Direct}), \texttt{+A} prompt, against the proposed prompt ensemble method (\texttt{Ensemble}). Performance is categorized into \textbf{hard-decision metrics} (Macro-F1, Micro-F1, and Top-1 Accuracy in \%) and \textbf{distribution-aware assessments}, including Kullback–Leibler Divergence (KLD), Jensen–Shannon Divergence (JSD), Total Variation Distance (TVD), Cosine Similarity (Sim $\times 100$), and Mean Squared Error (MSE). \textbf{\textcolor{red!85}{Bold red}} denotes the superior setting per corpus. $\uparrow$ ($\downarrow$) indicates that higher (lower) values are better.}
\vspace{-3mm}
\begin{tabular}{lll|ccc|ccccc}
\hline
\textbf{Dataset} & \textbf{Model} & \textbf{Prompt} & \multicolumn{3}{c|}{\textbf{Hard-decision Assessment}} & \multicolumn{5}{c}{\textbf{Distribution-aware Assessment}} \\
\cline{4-11}
 & & & \textbf{Ma-F1$\uparrow$} & \textbf{Mi-F1$\uparrow$} & \textbf{Acc$\uparrow$} & \textbf{KLD$\downarrow$} & \textbf{JSD$\downarrow$} & \textbf{TVD$\downarrow$} & \textbf{SIM$\uparrow$} & \textbf{MSE$\downarrow$} \\
\hline
\multirow{6}{*}{BIIC-Podcast (zh-tw, 8)} 
 & \multirow{3}{*}{Q2A} & Direct & 28.14 & 42.16 & 29.57 & 5.473 & 0.363 & 0.657 & 49.61 & 0.074 \\
 &  & +A & 4.79 & 9.38 & 9.67 & \textbf{\textcolor{red!85}{2.440}} & 0.402 & 0.749 & 45.78 & 0.073 \\
 &  & Ensemble & \textbf{\textcolor{red!85}{31.12}} & \textbf{\textcolor{red!85}{51.14}} & \textbf{\textcolor{red!85}{42.95}} & 7.257 & \textbf{\textcolor{red!85}{0.309}} & \textbf{\textcolor{red!85}{0.559}} & \textbf{\textcolor{red!85}{56.49}} & \textbf{\textcolor{red!85}{0.070}} \\
 \cline{2-11}
 & \multirow{3}{*}{AF3} & Direct & 15.57 & 25.70 & 22.95 & 4.767 & 0.374 & 0.680 & 46.97 & 0.083 \\
 &  & +A & 16.88 & 42.70 & 38.11 & 10.391 & 0.358 & 0.616 & 48.96 & 0.101 \\
 &  & Ensemble & 20.98 & 45.81 & 41.17 & 10.172 & 0.347 & 0.595 & 50.11 & 0.097 \\
\hline
\multirow{6}{*}{CREMA-D (en, 6)} 
 & \multirow{3}{*}{Q2A} & Direct & \textbf{\textcolor{red!85}{61.88}} & 65.34 & 49.03 & 4.112 & 0.196 & 0.443 & 69.43 & 0.054 \\
 &  & +A & 1.99 & 1.76 & 16.43 & \textbf{\textcolor{red!85}{1.092}} & 0.253 & 0.554 & 60.12 & 0.058 \\
 &  & Ensemble & 61.00 & \textbf{\textcolor{red!85}{68.25}} & \textbf{\textcolor{red!85}{59.16}} & 5.482 & \textbf{\textcolor{red!85}{0.175}} & \textbf{\textcolor{red!85}{0.389}} & \textbf{\textcolor{red!85}{75.81}} & \textbf{\textcolor{red!85}{0.045}} \\
 \cline{2-11}
 & \multirow{3}{*}{AF3} & Direct & 40.57 & 41.19 & 39.22 & 6.372 & 0.275 & 0.557 & 59.71 & 0.088 \\
 &  & +A & 48.59 & 54.14 & 53.34 & 6.024 & 0.239 & 0.501 & 65.67 & 0.073 \\
 &  & Ensemble & 54.06 & 63.98 & 57.60 & 6.652 & 0.209 & 0.446 & 71.34 & 0.064 \\
\hline
\multirow{6}{*}{EmotionTalk (zh, 7)} 
 & \multirow{3}{*}{Q2A} & Direct & 33.10 & 40.17 & 36.24 & 5.842 & 0.377 & 0.667 & 47.91 & 0.105 \\
 &  & +A & 8.95 & 13.36 & 21.51 & \textbf{\textcolor{red!85}{3.191}} & 0.428 & 0.770 & 41.55 & 0.111 \\
 &  & Ensemble & \textbf{\textcolor{red!85}{36.23}} & 47.11 & 47.33 & 6.607 & 0.323 & 0.583 & 55.13 & 0.091 \\
 \cline{2-11}
 & \multirow{3}{*}{AF3} & Direct & 23.80 & 31.41 & 31.99 & 3.599 & 0.369 & 0.661 & 49.19 & 0.100 \\
 &  & +A & 31.15 & \textbf{\textcolor{red!85}{54.39}} & 51.84 & 7.352 & 0.313 & 0.535 & 56.32 & 0.105 \\
 &  & Ensemble & 36.09 & 53.42 & \textbf{\textcolor{red!85}{54.07}} & 5.807 & \textbf{\textcolor{red!85}{0.282}} & \textbf{\textcolor{red!85}{0.499}} & \textbf{\textcolor{red!85}{60.37}} & \textbf{\textcolor{red!85}{0.087}} \\
\hline
\multirow{6}{*}{IEMOCAP (en, 4)} 
 & \multirow{3}{*}{Q2A} & Direct & \textbf{\textcolor{red!85}{62.83}} & \textbf{\textcolor{red!85}{62.10}} & \textbf{\textcolor{red!85}{60.03}} & 2.962 & \textbf{\textcolor{red!85}{0.234}} & 0.464 & \textbf{\textcolor{red!85}{69.01}} & \textbf{\textcolor{red!85}{0.121}} \\
 &  & +A & 1.30 & 1.32 & 20.53 & \textbf{\textcolor{red!85}{1.286}} & 0.338 & 0.680 & 54.53 & 0.158 \\
 &  & Ensemble & 56.00 & 59.24 & 54.23 & 5.369 & 0.254 & 0.472 & 63.66 & 0.152 \\
 \cline{2-11}
 & \multirow{3}{*}{AF3} & Direct & 50.95 & 55.30 & 51.89 & 4.716 & 0.261 & 0.465 & 62.14 & 0.160 \\
 &  & +A & 49.16 & 53.25 & 51.15 & 4.723 & 0.265 & 0.480 & 61.70 & 0.158 \\
 &  & Ensemble & 60.07 & 61.81 & 56.85 & 5.700 & 0.248 & \textbf{\textcolor{red!85}{0.449}} & 64.80 & 0.155 \\
\hline
\multirow{6}{*}{MSP-Podcast 2 (test1, 8)} 
 & \multirow{3}{*}{Q2A} & Direct & 24.43 & 34.72 & 18.85 & 10.371 & 0.421 & 0.715 & 37.14 & 0.087 \\
 &  & +A & 1.75 & 2.06 & 19.28 & \textbf{\textcolor{red!85}{1.741}} & 0.361 & 0.692 & 49.10 & \textbf{\textcolor{red!85}{0.060}} \\
 &  & Ensemble & \textbf{\textcolor{red!85}{33.76}} & 46.99 & 36.65 & 9.959 & 0.346 & 0.602 & 50.52 & 0.072 \\
 \cline{2-11}
 & \multirow{3}{*}{AF3} & Direct & 23.02 & 33.80 & 34.35 & 10.007 & 0.388 & 0.681 & 44.51 & 0.098 \\
 &  & +A & 24.11 & 36.36 & 33.33 & 7.541 & 0.359 & 0.656 & 49.50 & 0.082 \\
 &  & Ensemble & 30.18 & \textbf{\textcolor{red!85}{48.61}} & \textbf{\textcolor{red!85}{41.62}} & 10.415 & \textbf{\textcolor{red!85}{0.328}} & \textbf{\textcolor{red!85}{0.587}} & \textbf{\textcolor{red!85}{53.44}} & 0.078 \\
\hline
\multirow{6}{*}{MSP-Podcast 2 (test2, 8)} 
 & \multirow{3}{*}{Q2A} & Direct & 24.50 & 40.16 & 20.99 & 7.744 & 0.361 & 0.651 & 46.70 & 0.068 \\
 &  & +A & 1.33 & 2.17 & 5.84 & \textbf{\textcolor{red!85}{1.500}} & 0.334 & 0.649 & 52.35 & \textbf{\textcolor{red!85}{0.053}} \\
 &  & Ensemble & \textbf{\textcolor{red!85}{29.13}} & 48.04 & 43.68 & 9.292 & 0.313 & 0.558 & 57.11 & 0.059 \\
 \cline{2-11}
 & \multirow{3}{*}{AF3} & Direct & 17.98 & 34.76 & 39.80 & 8.283 & 0.321 & 0.593 & 55.39 & 0.074 \\
 &  & +A & 18.80 & 38.34 & 41.05 & 6.589 & 0.301 & 0.578 & 59.59 & 0.063 \\
 &  & Ensemble & 23.38 & \textbf{\textcolor{red!85}{48.65}} & \textbf{\textcolor{red!85}{55.70}} & 10.059 & \textbf{\textcolor{red!85}{0.284}} & \textbf{\textcolor{red!85}{0.521}} & \textbf{\textcolor{red!85}{63.11}} & 0.061 \\
\hline
\end{tabular}
\label{tab:soft_eval_comprehensive}
\vspace{-2mm}
\end{table*}

\begin{figure}[!h]
    \centering
    \includegraphics[width=0.44\textwidth]{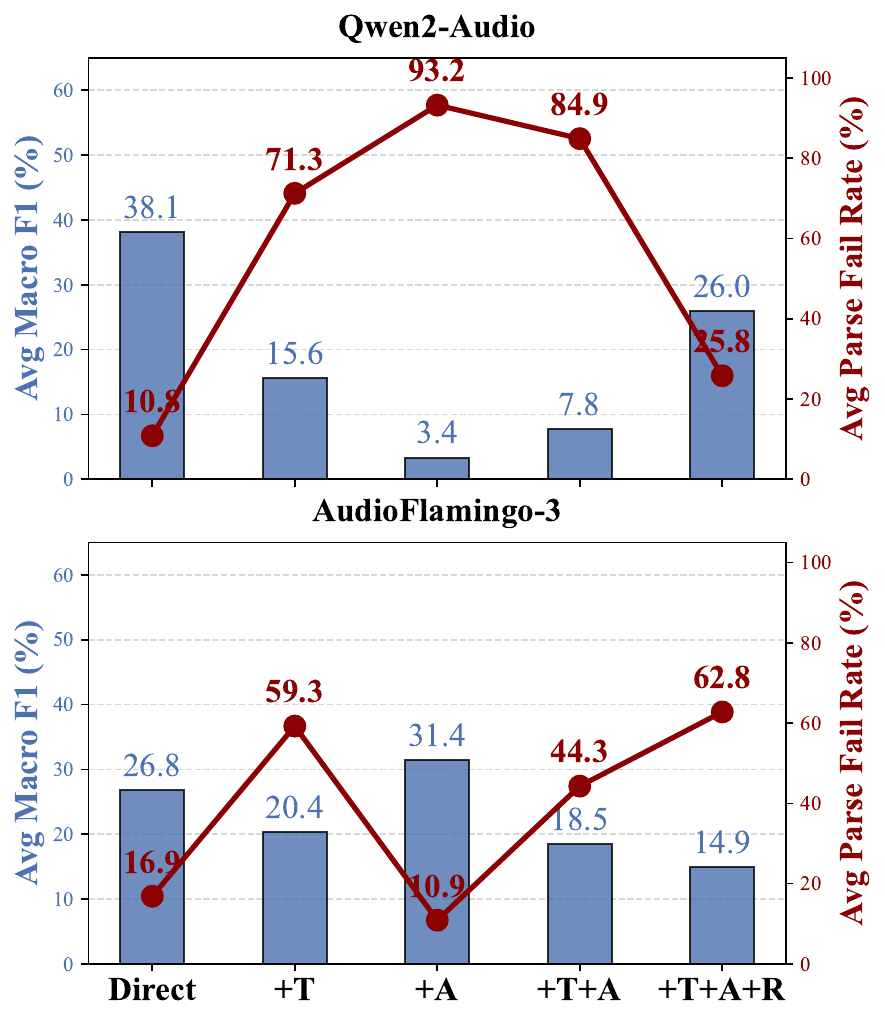} 
    \vspace{-3mm}
    \caption{Impact of zero-shot prompt complexity on performance and format adherence. The blue bars (left axis) denote the Average Macro-F1 across five evaluation corpora, while the red line (right axis) illustrates the Average Parse Failure Rate. As prompt constraints increase (from direct classification \texttt{Direct} to requesting intermediate acoustic captions \texttt{+T}), both Q2A and AF3 struggle to follow the structured text-output format, leading to a severe degradation in overall predictive performance. This zero-shot stochasticity necessitates our proposed distribution-aware prompt ensemble methodology.}
    \vspace{-4mm}
    \label{fig:prompt_trend}
\end{figure}

\vspace{-1mm}
\subsection{Soft-Label Evaluation}
\vspace{-1mm}
Table~\ref{tab:soft_eval_comprehensive} presents a comprehensive assessment of zero-shot soft-label performance across five corpora that provide multi-annotator distributions. 
Based on the prompt sensitivity observed in Figure~\ref{fig:prompt_trend}, we select three representative configurations for this detailed comparison: the baseline \textbf{Direct} prompt, the acoustic caption prompt (\textbf{+A}), and our proposed \textbf{Ensemble} strategy.

\vspace{-1mm}
\subsubsection{Impacts of Prompts and Zero-Shot Instability}
\vspace{-1mm}
A critical challenge in evaluating generative speech LLMs is inference-time stochasticity and strict adherence to structured output formats. 
As illustrated in Figure~\ref{fig:prompt_trend}, instructing models to perform intermediate reasoning tasks often disrupts their ability to generate valid JSON predictions. 
Interestingly, this degradation is highly model-dependent. 

For Q2A, increasing prompt complexity leads to a catastrophic surge in the Average Parse Failure Rate (red line)---peaking at 93.2\% for the \textbf{+A} prompt---which directly plunges the average Macro-F1 score to near zero.
Conversely, AF3 handles the \textbf{+A} prompt relatively well, achieving its highest single-prompt F1 (31.4\%) with a low parse failure rate (10.9\%). 
Nonetheless, both models suffer severe formatting failures when subjected to combined reasoning prompts (e.g., \textbf{+T+A+R}). 
This reveals that relying on a single prompt is inherently brittle and inconsistent across different LLMs.

\vspace{-1mm}
\subsubsection{Mitigating Instability via Prompt Ensemble}
\vspace{-1mm}
To mitigate this zero-shot instability, the proposed prompt ensemble strategy (\textbf{Ensemble}) is evaluated following \cite{zhang2026scaling}. 
This ensemble method aggregates discrete predictions across all five prompt variations using a uniform fallback mechanism for parse failures (as detailed in Section \ref{sec:zs-protocal}). 
As shown in Table~\ref{tab:soft_eval_comprehensive}, the ensemble method effectively neutralizes the formatting failures of individual prompts and consistently outperforms the baseline \textbf{Direct} prompt. 
Not only does it recover and improve hard-decision metrics (e.g., boosting Q2A's Macro-F1 from 61.88\% to 68.25\% on CREMA-D), but it also yields significantly better subjectivity alignment. 
Across most datasets, the \textbf{Ensemble} approach achieves higher SIM and lower JSD/TVD. 
This demonstrates that, rather than being a flaw, the stochasticity of diverse prompt responses can be aggregated to effectively approximate the uncertainty and subjectivity inherent in human emotion perception.

\vspace{-1mm}
\subsubsection{Qwen2-Audio vs. Audio Flamingo 3}
\vspace{-1mm}
Across the majority of the benchmarked corpora, Q2A consistently demonstrates superior alignment with human soft labels compared to AF3.  
This is particularly evident in the CREMA-D and IEMOCAP datasets, where Q2A achieves significantly higher Cosine Similarity ($\uparrow$) and lower MSE ($\downarrow$).  
These results suggest that Q2A's underlying representation is more effective at mimicking human annotators' consensus in structured environments.

However, one can observe a performance trade-off in naturalistic datasets such as EmotionTalk and BIIC-Podcast, where AF3 occasionally yields lower KLD and JSD despite having lower accuracy. 
This discrepancy indicates that while Q2A may be more accurate in top-1 prediction, its probability mass is often over-concentrated (over-confident), whereas AF3 maintains a smoother distribution that better reflects the inherent uncertainty in cross-lingual, spontaneous speech.


\begin{table}[!t]
\centering
\fontsize{7}{9}\selectfont
\caption{\small Comparison of best zero-shot speech LLMs against recent fully supervised state-of-the-art benchmarks on Hard-label Macro-F1 (\%).}
\vspace{-3mm}
\label{tab:sota_comparison}
\begin{tabular}{llc}
\hline
\textbf{Dataset} & \textbf{Model / Framework} & \textbf{Ma-F1} \\ \hline
\multirow{4}{*}{BIIC-Podcast (zh-tw, 8)} & \cite{wagner2023dawn} & 35.5 \\
 & \cite{Wu_2024} & 28.3 \\
 & \textbf{This paper (Q2A-Ensemble)} & 31.1 \\
 & \textbf{This paper (AF3-Ensemble)} & 21.0 \\ \hline
\multirow{5}{*}{CREMA-D (en, 6)} & \cite{chouMinorityViewsMatter2025} & 71.0 \\
 & \cite{wagner2023dawn} & 70.6 \\
 & \cite{Wu_2024} & 67.6 \\
 & \textbf{This paper (Q2A-Direct)} & 61.9 \\
 & \textbf{This paper (AF3-Ensemble)} & 54.1 \\ \hline
\end{tabular}%
\vspace{-4mm}
\end{table}

\subsubsection{Capturing Affective Ambiguity and Subjectivity}

To compare the performance of generative speech LLMs within the broader SER landscape, Table~\ref{tab:sota_comparison} compares the best zero-shot configurations against recent fully supervised state-of-the-art (SOTA) benchmarks, including conventional Transformers \cite{wagner2023dawn}, Self-Supervised Learning (SSL) representations \cite{Wu_2024}, and distribution-aware supervised models \cite{chouMinorityViewsMatter2025}. 
As shown in Table~\ref{tab:sota_comparison}, zero-shot speech LLMs' performance generally trails behind task-specific supervised models in hard-label Macro-F1. 
However, this performance gap underscores a fundamental difference in modeling paradigms rather than a mere capability deficit.  
Supervised baseline models are explicitly optimized via cross-entropy to collapse subjective annotations into hard labels, thereby maximizing categorical metrics such as F1. 
In contrast, zero-shot generative speech LLMs maintain a broader, untuned probability mass.

This brings us to a key contribution of the paper: the quantification of affective ambiguity.  
In naturalistic datasets such as the MSP-Podcast series, which are characterized by high inter-rater variability, the relatively low TVD and JSD scores for both Q2A and AF3 (as seen in 
Table~\ref{tab:soft_eval_comprehensive}) suggests that speech LLMs possess a strong latent capacity to model emotional subjectivity. 
Unlike traditional closed-set classifiers, which discard human perceptual variance, generative speech LLMs naturally represent competing emotional states in their output logits.  
The ability of these models to maintain high SIM to human label distributions without any dataset-specific fine-tuning demonstrates their exceptional potential for developing affect-aware systems that respect the diverse perceptions of multiple listeners.

\subsection{Cross-Corpus Evaluation}

Table~\ref{tab:cross_scenario} reports cross-domain transfer results for both models using each dataset's native label set. 
For Q2A, fine-tuning on a mismatched English source exceeds the best zero-shot prompt on 11 of the 12 source–target pairs. 
MELD is the most effective source, with a mean Macro-F1 gain of 17.7 over zero-shot across three targets. 
On IEMOCAP, it reaches 81.6, and on CREMA-D, 86.6. MSP-Podcast, the other in-the-wild corpus, also improves all three targets but with a notably smaller mean gain. 
Conversely, MELD is also the hardest target: the sole degradation in the table occurs when transferring IEMOCAP→MELD ($-$3.4), where IEMOCAP's 4 categories cover only 4 of MELD's 7, missing Disgust, Fear, and Surprise.

Under the same LoRA configuration, AF3 exceeds zero-shot on only 3 of 12 pairs, all targeting acted corpora with modest gains. 
All six transfers to in-the-wild targets (MELD, MSP-Podcast) degrade, consistent with AF3's in-domain SFT already falling below zero-shot on both corpora. MSP-Podcast is a particularly harmful source, accounting for the three largest drops in the table. 
As in Section~4.2, these results may partly reflect LoRA hyperparameter mismatch rather than a fundamental model limitation.

\begin{table}[!t]
\centering
\fontsize{7}{9}\selectfont
\setlength{\tabcolsep}{3pt}
\caption{\small Cross-scenario transfer (Macro-F1). Each cell shows the score when fine-tuning on the source (row) and evaluating on the target (column), with $\Delta$ over best zero-shot in parentheses. Diagonal (\textbf{bold}): in-domain SFT. MSP-Podcast uses test1.}
\vspace{-3mm}
\begin{tabular}{lllll}
\toprule
\textbf{Source$\backslash$Target} & \textbf{CREMA-D} & \textbf{IEMOCAP} & \textbf{MELD} & \textbf{MSP-Podcast 2} \\
\midrule
\multicolumn{5}{c}{\textit{Qwen2-Audio}} \\
\midrule
CREMA-D & \textbf{91.1} (+21.7) & 79.9 (+20.5) & 30.7 (+2.6) & 25.8 (+5.5) \\
IEMOCAP & 76.8 (+7.4) & \textbf{82.4} (+23.0) & 24.7 (-3.4) & 31.5 (+11.1) \\
MELD & 86.6 (+17.1) & 81.6 (+22.2) & \textbf{39.3} (+11.2) & 34.1 (+13.7) \\
MSP-Podcast 2 & 78.6 (+9.1) & 68.1 (+8.7) & 30.8 (+2.8) & \textbf{29.2} (+8.8) \\
\midrule
\multicolumn{5}{c}{\textit{AudioFlamingo-3}} \\
\midrule
CREMA-D & \textbf{75.2} (+15.4) & 65.1 (+1.1) & 21.6 (-8.2) & 22.9 (-5.7) \\
IEMOCAP & 63.6 (+3.8) & \textbf{65.2} (+1.2) & 20.3 (-9.5) & 19.2 (-9.5) \\
MELD & 63.3 (+3.5) & 62.0 (-2.0) & \textbf{25.7} (-4.1) & 23.5 (-5.2) \\
MSP-Podcast 2 & 36.6 (-23.2) & 45.4 (-18.6) & 18.9 (-10.9) & \textbf{19.1} (-9.6) \\
\bottomrule
\end{tabular}
\label{tab:cross_scenario}
\vspace{-6mm}
\end{table}

\section{Conclusion, Limitation and Future Work}
This paper introduces VoxEmo, an evaluation toolkit and benchmark that standardises inference-time protocols for speech-LLM-based SER across 35 corpora in 15 languages. 
Zero-shot performance is highly sensitive to prompt design, and the acted–naturalistic split in corpus construction systematically shapes prompt effectiveness. 
Supervised fine-tuning narrows the gap with traditional baselines but does not close it in most cases, with effectiveness depending on dataset scale and the choice of foundation model. 
Nevertheless, even without task-specific training, zero-shot outputs capture affective ambiguity that aligns with human annotation distributions, and the generative interface enables cross-domain transfer across mismatched label sets—though both benefits vary sharply between models.

Several limitations should be noted. 
The benchmark evaluates only two models that share the same audio encoder (Whisper-large-v3) and parameter scale; whether the observed patterns generalise to models at larger scales or with different audio front-ends, including recent tokenizer-based architectures, remains open. Such tokenizer-based models cannot currently be served in the same inference environment, limiting the direct comparison. 
A single LoRA configuration is used throughout, and the limited SFT gains of AF3 may partly reflect hyperparameter mismatch. 
Soft-label evaluation is restricted to the five corpora that provide per-annotator metadata. 
Additionally, this work focuses on corpus-level aggregate metrics and does not examine within-dataset factors such as class imbalance, per-class performance, or speaker effects. 
Future work will extend the benchmark to architecturally diverse models, investigate SFT hyperparameter sensitivity, and incorporate finer-grained per-dataset analyses.

\section{Generative AI Use Disclosure}

Generative AI tools were employed solely to improve grammar, clarity, and readability. 
The authors are fully responsible for all technical content and results presented in this paper. 
These tools were not used to generate scientific ideas, analyses, or findings.

\bibliographystyle{IEEEtran}
\bibliography{mybib}

\end{document}